# Unaccounted source of systematic errors in measurements of the Newtonian gravitational constant G.

Riccardo DeSalvo [1,2]

[1] California State University, Northridge, 18111 Nordhoff Street, Northridge, CA 91330-8332 (USA).
[2] University of Sannio, Corso Garibaldi 107, Benevento 82100 (Italy).
e-mail address: Riccardo.desalvo@gmail.com

Many precision measurements of G have produced a spread of results incompatible with measurement errors. Clearly an unknown source of systematic errors is at work. It is proposed here that most of the discrepancies derive from subtle deviations from Hooke's law, caused by avalanches of entangled dislocations. The idea is supported by deviations from linearity reported by experimenters measuring G, similar to what observed, on a larger scale, in low-frequency spring oscillators. Some mitigating experimental apparatus modifications are suggested.

## I. INTRODUCTION

Disturbing inconsistencies have been reported in the measurements of G [1, 2, 3, 4, 5, 6, 7]. The data scatter is about an order of magnitude larger than the calculated measurement errors. The likely common point of failure is that most of the measurements were made with torsion balances suspended by Tungsten wires or ribbons, or involve metal flexures of various kinds. The precision of these experiments is based on the strict validity of Hooke's law, i.e. the elastic material produces a restoring force proportional to displacement, which tends to return the system to a fixed equilibrium position. A new, low-frequency dissipation mechanism was recently identified, which explained many deviations from linearity and predictability observed in seismic isolation springs for Gravitational Wave detectors [8]. The mechanism, caused by avalanches of entangled dislocations, is observable only below a low frequency threshold, but then it produces subtle deviations from Hooke's law. From the point of view of G measurements, the most damaging effect is an unceasing and intrinsically stochastic shift of the equilibrium point affecting fibers, ribbons and flexures. This mechanism, that also produces the well-known and well-modeled hysteresis, is common to polycrystalline metals, but was not taken into account in measurements of G. The deviations from linearity that this model entails offer a single and simple explanation for many discrepancies. Importantly, unexplained behaviors reported by G-measurement experimenters are indicative that a process of this kind has been at work. The same mechanism can be expected to affect tests of the weak principle of equivalence and searches for violations of the inverse square law of gravity as well as any other precision measurement that involve metal torsion fibers, ribbons, or flexures.

### 1. Early corrections and failure of Elasticity

Deviations from Hooke's law were already identified as sources of systematic biases in G measurements. Initially loss mechanisms were modeled with viscous damping, i.e. energy losses proportional to oscillation speed. In 1992 Quinn et al. [9] found significant deviations from Hooke's law, and frequency independent losses, that forced them to consider a continuum of dashpots. They suggested *"a hierarchical regime in which successive levels of dislocation unlocking occur",* a wording that comes close to some of the conclusions of this paper. In 1995 Kuroda [10] discussed the changes of the measured value of G when the dissipation in the suspension is not viscous. His improved model used a phase-delay model to describe the dissipation mechanism, i.e. included a frequency independent imaginary component to the elastic constant of the suspension wire of a torsion balance. His corrections affected most of the previous G measurements, whose results were changed accordingly. The phase-delay model of the loss mechanism was an important improvement over the viscous one, but, in absence of an appropriate microscopic theoretical understanding, it could not take into account some key details and in many ways proved unsatisfactory. Other unexplained and potentially dangerous deviations from linearity were observed: Matsumura [11] and Yang [12] found frequency-dependent Young's modulus variations in Tungsten wires and Hu [13] an amplitude dependency. Both effects have also been observed in springs of gravitational wave detectors. The above observations were in part made in support of Kuroda's model, but with present knowledge can be reinterpreted in view of new the model discussed here.

## 2. A different theoretical framework

Metals are naturally plastic, due to the free movement of dislocations, filament-like defects that extend across the entire width of grains. In pure metals, dislocations move more or less freely in response to changing stresses and carry deformation. Hundreds of thousands of km of active dislocations are present in every cubic cm of polycrystalline metals. Elastic materials are obtained by the addition of point-like impurities or precipitates, which anchor the deformation-carrying dislocations [14]. However, only a limited number of impurities can be added and residual deviations from elasticity remain.

Seismic isolation for gravitational wave detectors needs to be pushed to the $10^{-20}$ m/√Hz level. Both materials and methods used have been studied in depth for more than 15 years. Three kinds of mechanical filters have been used, the simple and the inverted pendulums for horizontal attenuation, and cantilever springs for the vertical direction [15]. Maraging steel, a high quality and very elastic material proposed by Renzo Valentini and by the author, is now almost universally adopted for seismic attenuation in gravitational wave detectors. The choice was driven by the fact that Maraging steel has a yield point above 1.8 GPa and, in proper load conditions, it is essentially creep free [16]. Significant deviations from Hooke's law and from both viscosity and hysteresis modeled dissipation mechanisms were observed in all three kinds of mechanical filters, in Maraging steel as well as in different metals. Mechanical losses of Maraging steel were extensively studied using 40 cm long triangular blades in the monolithic geometric anti-spring filter configuration [17], which is an almost ideal investigation tool because it cancels the spring's elastic constant and exposes the material's non-linearities. The blades were loaded to ~1 GPa surface stress, i.e. well below the material's 1.8 GPa yield point. Careful thermal treatments, including 100-hour precipitation at 435°C and post-loading thermal anneal cycles, were applied to eliminate any significant creep. Clamping problems were ruled out by partial loosening of the clamps and finding no worsening of the key discrepancies. Eventually the discrepancies were explained in terms of entanglement and disentanglement of dislocations inside the metal grains, a process controlled by Self Organized Criticality (SOC) statistics [18]. Dislocation SOC is an extension of the Granato-Lueck theory, made necessary by the fact that dislocations, which are line defects that cannot cross each other without large expenditure of energy, are more likely to be entangled with other dislocations than locked by point-like defects. Thus, while point-like defects play a crucial role in pinning dislocations, entanglement is even more important.

It has been observed that entanglement is far from complete and stable. Changes of stress induce cascades of disentanglement during which the dislocations descend the stress field and re-entangle in a different location. This process causes metastable changes of shape, i.e. of the equilibrium position of polycrystalline metal flexures. My collaborators and I reported in detail several deviations from pure elasticity of the same types reported in Tungsten wires. Therefore, I suggest that while the phase delay model adopted by Kuroda mimics very closely the behavior of materials, it is not complete. The detailed effects of dislocation SOC statistics are intrinsically non-predictable, due to the avalanche character of disentanglement/re-entanglement events. The behavior results from the history-dependent entanglements inside the materials, which further complicate predictions. Paper [8] contains numerous figures illustrating the points discussed here. In this paper, when necessary for purposes of reference, those figures will be labeled with an asterisk. Those measurements were performed in Maraging steel, but the theoretical model applies to all polycrystalline metals. Besides Tungsten wires, numerous instances of behavior compatible with collective dislocation activity were observed in deeply drawn, high-carbon piano wire [19], Nispan-C [20] and in Copper-Beryllium flexures [9, 21, 22].

A parallel can be found in Maraging steel for virtually all of the perplexing behaviors observed in Tungsten fibers, some of which are listed in the next chapter. The Young's modulus and dissipation were shown to be amplitude dependent at low frequency (FIG. 20*, 21* and 24* for the modulus and FIG. 23* for dissipation), and a sizable amount of anelasticity (FIG. 9* and 12*) is present. It was shown that the hysteresis amplitude is strongly dependent on the oscillator resonant frequency (FIG. 8*, 10*, 11* and 14*).

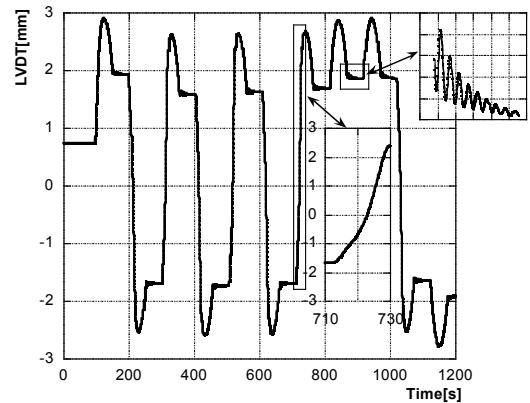

FIG. 1. Adapted from [8]. Hysteretic response to pulsed, half-sinusoid-shape excitations of 26 mN peak amplitude (~35 ppm of the static load), applied with a voice coil; the pulses were alternating sign and then same sign to separately probe hysteresis and anelasticity. The initial response to a changed-sign excitation is disproportionate and causes large hysteresis. The left insert focuses on the large drifting of the equilibrium point, which is followed by a more normal response to the excitation and the anelastic response. Note that the runoff of the equilibrium position

has initially the wrong curvature (concave), differing the shape of the excitation force is always convex. This is due to triggering of extensive, merged, dislocation avalanches. Runoff and the wrong curvature response are only present when the excitation pulse sign is reversed and not if excitations of the same sign are applied (T >700 s). The history dependent material's behavior is evident from the pulse-to-pulse response variations. The right insert shows the anelastic behavior at the end of an excitation pulse. The lifetime of the anelastic decay, fitted with an exponential, is of 15.8±0.2 s.

In Maraging steel the anomalous hysteresis due to the disentanglement/re-entanglement process has an onset characteristic time of seconds as illustrated in FIG. 1 in this paper and FIG. 27*. This is in addition to an anelastic relaxation mode with time constant longer than ten seconds (insert FIG. 1 and FIG. 9* and 12*). Indeed, although an exponential function was successfully used in the fit, anelasticity may have no specific exponential time constant and appear to be made of infinite time constants, as suggested by [23]. The lack of time constants is specific to SOC dominated processes.

An increase in hysteresis of more than one order of magnitude was observed in Maraging steel when exploring lower frequencies, in addition to the anelasticity. The discrepancies in G measurements are of the order of 1,000 ppm, 10,000 times smaller than the effects observed in Maraging steel springs. It is clear that, although the materials chosen for the G measurements are less lossy than Maraging steel, and the ill effects much more diluted, the frequency and amplitude dependent hysteresis and anelasticity documented in [8] have the potential to explain all discrepancies in G measurements. It is worth noting that the transition from "viscous" to SOC controlled distribution was observed in Maraging steel with characteristic time constants of a few seconds. The higher binding energy (melting point) of Tungsten may induce the transition at lower frequency, perhaps in or near the frequency range of torsion pendula. The behavior depends on the metallurgical and operational history of each specific sample. The intrinsically non-fully-predictable character of SOC-induced hysteresis makes matters even worse. It is easy to see why an a posteriori, Kuroda-style, correction become practically impossible. It should be noted here that frequency independent losses were observed in several materials as early as 1927 [24]. Quinn et al. found a similarly frequency independent losses in Copper-Beryllium [9] even connecting their model to a SOC-like 1/f noise in materials. They likewise found in 1995 [21, 22] that Copper-Beryllium modulus defect is independent of stress, as long as the stress is below creep level, and discussed the amplitude-dependent damping, behaviors that are easily explained by an entangled dislocation model. Because of the limitations of the visual and chart recorder acquisition techniques used at the time, they partially missed the importance of the metastable states produced by the SOC behavior of dislocations. One cannot avoid thinking that, if they had knowledge of the SOC theory, which was being invented in those years, and access to modern acquisition techniques, they would have reached conclusions similar to the ones presented here. Cagnoli et al. in 1993 [25] raised the possibility that SOC of dislocations controls dissipation in metals, without considering its effects on G-measurements.

### 3. Effects on G measurements

Deviations from Hooke's law found in gravitational wave isolation systems were studied in geometries very different from the torsion pendula used in G and principle of equivalence measurements, but similar deviations were found in torsion pendula or flexures. The effects of any equilibrium point instability in Cavendish-type measurements are obvious. Measurements based on frequency shifts were developed as an alternative to angular shift methods. They may appear to be less prone to SOC-induced problems, but some important caveats apply. If the equilibrium point shifts during each pendulum oscillation, the formulas used by Kuroda to calculate the corrections for the time of swing method become just an approximation of a more complex problem. In addition, large, amplitude-dependent changes of Young's modulus, presumably mimicked by equilibrium point dynamic shifts, were observed in FIG. 22*. Large amplitude and frequency dependent changes in loss angle were observed as well (FIG. 23*). These effects directly influence time of swing methods, and there is historical evidence that this might have been the case. Matsumura reported a frequency dependent Young's Modulus [11], Hu reported it to be amplitude dependent, [13], Yang [12, 26] reported a somewhat mysterious exponential Tungsten wire "aging" at the beginning of each measurement, which could be interpreted as a slow settling of the dislocation landscape disturbed by the re-excitation process. Significantly, wires in most settings presented long term drift, lasting months, which is compatible with entangled dislocation relaxation processes. Heyl and Chrozanovski in 1942 [27] used annealed and hard drawn Tungsten fibers, which gave different results for G (6.6685±0.0016 and 6.6755±0.0008 respectively). The two differing results are a smoking gun that dislocation activity influences time of swing G measurement results. Note that if dislocation SOC activity introduces a bias into the measurement of G, the averaging between the two results done by the authors may not be appropriate: the correct value would be on one of the sides, not in between. Considering dislocation SOC effects, it is not surprising that the annealed fiber, in which dislocation entanglement is reduced and dislocations have more phase space to move around and entangle, would produce more measurement spread than the hard drawn case.

Recent torsion balance measurements were made with ribbons to take advantage of gravitational restoring torque, and dilute the effects of variations of Young's modulus discussed by Kuroda [28, 29]. Yet, a few percent of the ribbon restoring force is still elastic. Therefore ribbons are subject to systematic errors from dislocation SOC noise. Feedback controls were introduced to improve G-measurements by nulling the angular motion [30, 28, 31, 29]. It would appear that a null oscillation method may be exempt from equilibrium point dragging, however FIG. 4* and 5* show that even if feedback nulls the motion, the internal forces that cause changes of equilibrium position appear as changes of required feedback amplitude, and therefore as signal. A suppressed oscillation method can be effective only if the angular motion is completely impeded and if oscillation is the only source of SOC activation. Pendulum oscillations or vibrations can be independent sources of dislocation disentanglement. In addition, the reduction of oscillation is dependent on the gain applied, which is likely held at reasonably low values to avoid noise injection. The residual motion, and there is always some, may easily generate non-negligible equilibrium point dragging effects.

Non torsional pendulums were tried. Gravitationally-induced pendulum deflections were used to measure G. Pendulums are affected by wire hysteresis: there is a dilution factor because most of the restoring force is gravitational, 87% in [32] using Tungsten wires and only a fraction is due to elasticity, but *"these springs are stiff compared to a torsion fiber"*. The effects of equilibrium point drag already proved quite notable in other suspension wires, as reported by Greenhalgh [19]. The potential for mischief from hysteresis is significant enough to consider re-evaluation of systematics in terms of dislocation SOC noise at the flexing point.

The beam balance measurement of [33] is a completely different system, immune to the problems of flexing or torqueing metal wires. In addition, the careful alternating weighting of two test masses after each movement of the field masses is designed to cancel systematic errors. However the balance they used has flexures, which may be subject to SOC noise: the authors report mysterious rapidly growing and then slowly decreasing zero-point variations after switching conditions, which may be due to relaxations induced inside the balance flexures by load changes. It is not clear if, and how much, these changes, assuming that they are dislocation-SOC-induced, could affect the overall measurement of G. Only the authors can solve this fine point, revisiting their careful systematic evaluation in view of the peculiarities of the SOC noise source.

## 4. Mitigating methods and solutions

Theoretically, one could think of calculating correction factors for existing measurements. Knowledge of the hysteresis for each specific wire at different amplitudes and different frequencies, as well as other wire-specific quantities, would be necessary for this evaluation. It is therefore impossible for an external person to attempt to estimate the corrections.

Modified measurements involving forced, damped oscillations, similar to the watch demagnetization processes, may bring some improvement to torsion pendulum measurements. The decreasing oscillations tend to smooth out the dislocation landscape, and bring the system far from critical slopes (FIG. 14* and [34]). But forced oscillations do not completely eliminate all of the dislocation SOC effects.

Bantel and Newman mitigate dissipation by cryogenic cooling [35,36].

SOC activity exists only in the presence of a driving source. For example, the avalanches on the steep side of dunes are caused by the wind that pushes over the top one grain of sand at the time. In the case of suspension wires or ribbons, in addition to the torsion pendulum motion itself, the excitation of dislocation activity may come from thermal excitation, vibrations, or other sources. The random walk in the Virgo inverted pendulums may be driven by ground vibrations, which are larger during storms. Careful shielding of torsion balances from seismic vibrations, noise and thermal drifts will help, but is not a guarantee.

Assuming, *or to prove that*, dislocation SOC is the source of the G-measurement discrepancies, the only radical solution in torsion pendula is to use *glassy fibers, or other elastic material without dislocations*. Fused silica fibers already proved to be almost ideal for precision measurements; they are the suspension material of choice for Gravitational Wave detectors and other precision experiments. Unfortunately, fused Silica, as discussed by Yang [12], is an electrical insulator and electrostatic effects on the torsion pendulum bob were found to be worse than the cure. Luther and Bagley [37, 38] showed that metal-coated fibers are not useful either. A proposed compromise solution, as suggested by Yang, may be using doped fused silica with a concurrent sacrifice in Q-factor.

Glassy metals do not have dislocations, have high Quality factors and are electrically conductive, thus keeping the payload grounded at all times. They have a yield point roughly twice that in a polycrystalline state, and can carry the same load on a thinner fiber, thus reducing the angular restoring force, increasing the signal and halving any residual systematic errors. They may be the ideal and offer a relatively easy solution to improve the G machines. Glassy metal fibers can be fabricated pulling glassy metal rods with an apparatus similar to the one developed by GEO [39] to pull fused silica mirror suspension wires. The only change would be to use an RF coil, instead of a laser, as the heat source. An additional advantage of pulling the fiber from a rod is that it leaves two natural handles that are easy to fasten using precision commercial collet chucks.

Glassy metal ribbons are another option. Ribbons a few µm thick are routinely spun-cast for use in high frequency transformers. Non-magnetic glassy metals can be spun as well.

Of course purely gravitational falling atom interferometry methods [40], and fiberless systems like the floating test mass system used by Michaelis [41, 42] are also immune to systematic errors from metal wire elasticity noise.

## II. CONCLUSIONS

Dissipation in torsion pendulum metal wires has been described with viscous, loss-angle and stick-and-slip mechanisms. The theory of Self Organized Criticality of entangled Dislocations offered an explanation for the many deviations from linearity and for the 1/f noise of springs and wires used in seismic attenuators for Gravitational Wave Detectors. It was argued in this paper that the same theory of entangled dislocations unifies the three dissipation models previously used and can explain the discrepancies between different G measurements. Given the emergent and history-dependent behavior of SOC systems, Kuroda-style corrections for this effect are not possible. Simple changes of the G-machines have been proposed that may disprove this supposition, or finally generate coherent evaluations of the value of G.

## ACKNOWLEDGEMENTS

I would like to thank Kazuaki Kuroda for putting me on this track, and Sydney Meshkov for intimating to me to write this article.